# Nanocrystalline superconducting $\gamma$-Mo$_2$N ultra-thin films for single photon detectors

J. A. Hofer,[1] M. Ginzburg,[1] S. Bengio,[2] N. Haberkorn,[1,2]

[1] Instituto Balseiro, Universidad Nacional de Cuyo and Comisión Nacional de Energía Atómica, Av. Bustillo 9500, 8400 San Carlos de Bariloche, Argentina.

[2] Comisión Nacional de Energía Atómica and Consejo Nacional de Investigaciones Científicas y Técnicas, Centro Atómico Bariloche, Av. Bustillo 9500, 8400 San Carlos de Bariloche, Argentina.

We analyze the influence of the surface passivation produced by oxides on the superconducting properties of $\gamma$-Mo$_2$N ultra-thin films. The superconducting critical temperature of thin films grown directly on Si (100) with those using a buffer and a capping layer of AlN are compared. The results show that the cover layer avoids the presence of surface oxides, maximizing the superconducting critical temperature for films with thicknesses of a few nanometers. We characterize the flux-flow instability measuring current-voltage curves in a 6.4 nm thick Mo$_2$N film with a superconducting critical temperature of 6.4 K. The data is analyzed using the Larkin and Ovchinnikov model. Considering self-heating effects due to finite heat removal from the substrate, we determine a fast quasiparticle relaxation time $\approx$ 45 ps. This value is promising for its applications in single-photon detectors.

*e-mail*: nhaberk@cab.cnea.gov.ar.

## 1. Introduction

The superconducting properties of transition metal nitrides (TMNs) have become an area of growing interest. Nitrides of niobium, titanium, zirconium and molybdenum have potential applications in electronic devices such as superconducting nanowire single-photon detectors (SNSPD) [1,2]. These devices consist of a micropatterned wire biased by a dc current slightly below the critical value $I_c$. When a photon interacts with the superconducting path, it produces a resistive "hotspot" that disrupts the superconductivity and generates a voltage pulse. After detection, the device recovers the zero resistance state after a characteristic relaxation time of quasiparticles ($\tau_e$). The latter relates to the time in which excited quasiparticles recombine into Cooper pairs through electron-electron and electro-phonon scattering [1].

Materials usually employed to design SNSPD are amorphous superconductors such as WSi [3] and MoSi [4,5] and crystalline NbN [6]. These systems present uniform superconductivity in films with thicknesses of a few nanometers and display a $\tau_e$ shorter than the ns. An alternative material for applications in detectors of radiation is $\gamma$-$Mo_2N$ [7,8]. Nanocrystalline $\gamma$-$Mo_2N$ preserves superconductivity in ultra-thin films [9,10]. The superconducting critical temperature ($T_c$) in thick films increases from $\approx$ 5 K to 8 K by disorder at the nanoscale and reduces for rich nitrogen stoichiometry [11,12,13,14]. Despite these interesting qualities, a particularity of molybdenum nitride thin films that can affect ultra-thin films' potential applications is the surface degradation by oxidation [9,12]. The presence of oxides at the interfaces is expected to reduce the section and suppress the superconducting properties of the nanowires.

In this work, we analyze the $T_c$ as a function of the thickness in ultra-thin films of $\gamma$-$Mo_2N$ grown by reactive sputtering on Si (100). To analyze the surface passivation's role, we compare the $T_c$ values of thin films growth directly on the substrate with those grown in a sandwich structure with a 1 nm thick AlN buffer/capping layer. The AlN was selected considering its high stability at the interface with nitrides [15]. The presence of oxides was studied by X-ray photoelectron spectroscopy (XPS). The results show that sandwich structures avoid the formation of oxides and maximize the $T_c$ in ultra-thin films. Moreover, using voltage-current curves and the instability in the flux vortex motion [16], we analyze the $\tau_e$ value for optimized sandwich structures with thicknesses of 6.6 nm and $T_c$ of 6.4 K. The $\tau_e$ is compared with those previously reported for materials usually applied in SNSPD.

## 2. Materials and methods

Molybdenum nitride thin films were grown by reactive sputtering on (100) Si as described in detail elsewhere [9,12]. No intentional heating of the substrate was used. To analyze the influence of dead interfaces, we design sandwich structures with a buffer/capping layer of aluminum nitride. The target power was RF 100 W (AlN) and DC 50 W ($Mo_2N$) and the total pressure at the chamber was 5 mTorr (93:7, Ar: $N_2$). During deposition, the substrate (typical size 1 $cm^2$) is positioned directly over the targets at ~ 5.5 cm. The residual pressure of the chamber was less than $1 \times 10^{-6}$ Torr. The thicknesses of the buffer and capping layers were fixed in $\approx$ 1 nm. The growth rates are $\approx$ 2 nm / min for $Mo_2N$ and $\approx$ 8 nm / min for AlN. Wherever used, the $[Mo_2N]_d$ and $[Mo_2N/AlN]_d$ notations indicate $Mo_2N$ films with thickness $d$ without and with a sandwich structure, respectively. The $d$ value in the $Mo_2N$ layer was varied between $\approx$ 3 nm and $\approx$ 6.4 nm.

The thicknesses of the $Mo_2N$ layers were estimated from low angle X-ray reflectivity (XRR) measurements using a Panalytical Empyrean equipment. Profile fitting was done using the Parratt32 code [17]. XPS was performed in high-vacuum conditions using a standard Al/Mg twin-anode, X-ray gun and a hemispherical electrostatic electron energy analyzer. All XPS spectra were calibrated to the C 1s peak at a binding energy of 284.8 eV. The electrical transport measurements were performed on 80 $\mu$m (length, $L$) x 6.8 µm (width, $w$) (Bridge 1, B1) and 80 µm (length, $L$) x 7.2 µm (width, $w$) (Bridge 2, B2) of $[Mo_2N/AlN]_{6.4}$ using the standard four-terminal transport technique. The bridges were fabricated using optical lithography and argon ion milling. The characteristic current-voltage (IV) curves were obtained with a Keithley Nanovoltmeter Model 2128A and a Keithley Current source Model 6221 AC/DC operating in synchronized mode with pulse duration of 0.2 ms.

### 3. Results and discussion

We analyzed the film surface chemical composition and the Mo oxidation state with and without a capping layer of AlN by XPS. As we have shown before [9,12], pristine films usually display oxides. Figures 1$a$ and 1$b$ display a comparison between Mo3d spectra of thin films with and without a 1 nm thick AlN protecting layer. The Mo3d spectra were fitted using a Voight function for $MoO_3$ and $MoO_2$ and a Doniach-Sunjic function for $Mo_2N$ with an asymmetry parameter of 0.15, plus a Shirley-type background. Three components were identified in the $[Mo_2N]_5$ surface. The major component at binding energy BE ≈ 228.5 eV can be assigned to $Mo_2N$ [18], a minor component at BE ≈ 230 eV can be assigned to $MoO_2$ [19], and an intermediate intensity component at BE ≈ 232.7 can be assigned to $MoO_3$ [20]. On the other hand, the spectra for $[Mo_2N/AlN]_5$ only display the component expected for $Mo_2N$, which clearly indicates that the oxides produced by atmospheric humidity in $[Mo_2N]_5$ are prevented by the AlN capping layer. The presence of oxides with a thickness of around 1 nm should directly impact the superconducting properties of ultra-thin films by reducing the effective thickness. Moreover, as we have shown before, interstitial oxygen suppresses the superconducting critical temperature compared to pure nitride films [21].

Figure 2$a$ and 2$b$ show typical XRR data and the corresponding fits for $[Mo_2N]_d$ ($d$ = 3.8 nm and 5.1 nm), and $[Mo_2N/AlN]_d$ ($d$ = 2.7 nm, 3.8 nm, 4.6 and 6.4 nm), respectively. The fits using Parrat32 reproduce well the data in all the cases. The film roughness is around 0.2 nm, which agrees with previous data by atomic force microscopy [9]. The data for $[Mo_2N]_d$, were fitted using a bilayer composed of a $Mo_2N$ film and a thin oxide capping layer. The results show that the best fits were obtained considering a reduction of around 1 nm in the effective thickness of the nitride and an extra layer. For sandwich structures, the data is consistent with a buffer and a capping AlN with a thickness of ≈ 1 nm. Figure 2$c$ shows the temperature dependence of the normalized resistance for the different films. The summary of $T_c$ is displayed in Fig. 2$d$. All the films show a $T_c$ suppressed with respect to the expectations for bulk (≈ 7.6 K) [12] with values that range from ≈ 3 K to 6.4 K. The main difference between single films and sandwich structures is suppression of $T_c$ in the former. The reduction in $T_c$ agrees with a lower effective thickness, as is expected from the presence of oxides. It is important to note that the interface between silicon / $Mo_2N$ also may produce an ultra-thin dead layer that is avoided in a sandwich structure with AlN.
To analyze the superconducting properties of the films in more detail, we performed measurements of the upper critical field ($H_{c2}$) with the magnetic field $B$ applied parallel and perpendicular to the surface (S). The analysis is conducted in the film with the largest $T_c$ and a thickness of 6.4 nm (sandwich structure). The data with **B** $\perp$ S provides information about the coherence length $\xi$. On the other hand, the **B** // S is expected to present a 2D behavior related to the $\xi$ / $d$ ratio. Figure 3 displays the results for **B** $\perp$ S and **B** // S. The data for **B** $\perp$ S was analyzed by the Werthamer-Helfand-Hohenberg (WHH) model developed for dirty one-band superconductors [22], where:

$$ln\frac{1}{t} = \sum_{v=-\infty}^{\infty}\left(\frac{1}{|2v+1|} - \left[|2v+1| + \frac{\hbar}{t} + \frac{(\alpha\hbar/t)^2}{|2v+1|+(\hbar+\lambda_{so})/t}\right]^{-1}\right), \qquad [1]$$

with $t = T / T_c$, $\hbar = (4/\pi^2)(H_{c2}(T)/|dH_{c2}/dT|_{T_c})$, $\alpha$ is the Maki parameter which quantifies the weakening influence of the Pauli electron spin paramagnetism on the superconducting state, and $\lambda_{so}$ is the spin-orbit scattering constant. The WHH formula satisfies the relation $H_{c2}(0) = \frac{H_{c2}^{orb}(0)}{\sqrt{1+\alpha^2}}$ when $\lambda_{so} = 0$ [23]. The $H_{c2}(T)$ curve adjusts to this model considering $\alpha = 0$, $\lambda_{so} = 0$ and $-|dH_{c2}/dT|_{T_c} \approx 3.3$ T/K. The $H_{c2}(0)$ obtained from the extrapolation to zero field is 10.5 T, which using $\xi_{GL}(0) = \sqrt{\Phi_0/(2\pi H_{c2}^{\perp}(0)}$ gives $\xi(0) = 5.5$ nm. The quasiparticle (electron) diffusion constant $D$ for the film can be estimated as $D = 4k_B/(\pi e d B_{c2}/dT)$ [24], being $\approx 0.33$ cm²/s.

Considering that the film has a thickness lower than 4.4$\xi$, the $H_{c2}(T)$ with **B** // S can be analyzed in the 2D limit as:

$$H_{c2}^{\parallel S}(T) = \frac{\sqrt{3}\Phi_0}{\pi d \, 0.855\xi(0)}(1 - T/T_c)^{1/2}. \qquad [2]$$

The equation considers that in the dirty limit $\xi(T) = 0.855\xi(0)\left(1 - T/T_c\right)^{-1/2}$. The data is well fitted with $d = 6.8$ nm, which is slightly thicker than the 6.4 nm obtained from XRR.

The Mo₂N films display weak pinning with very low critical current densities ($J_c$), and no features related to correlated disorder are observed [9]. The pearl lengths $\Lambda = \lambda^2/d$, relevant for screening of the magnetic flux in B1 and B2 using a penetration depth $\lambda$ (0) $\approx$ 800 nm [9], are estimated at $\approx$ 90 $\mu$m (more extensive than the bridge width). Next, we analyze the vortex pinning properties and Larkin and Ovchinnikov (LO) instability measuring IV curves as a function of $B$ for [Mo₂N/AlN]₆.₄. The measurements were performed with **B** $\perp$ S. The vortex pinning properties are determined from the magnetic field dependence of the critical current densities ($J_c$) at low magnetic fields applying currents to reach the flux-flow regime. The LO instability is analyzed in a broader range of fields applying currents up to the normal state.

To determine the magnitude of $J_c$ on B1 and B2, we use the criterion of 100 $\mu$V. Typical IV curves at self-field (SF) and under applied field at 3 K are depicted in the inset of Fig. 4. The IV curve at SF displays an abrupt jump at $J_c$. In contrast, those obtained applying magnetic display a zero resistance state at low currents and dissipation above the critical current ($I_c$) due to vortex motion. The flux-flow state manifests as a linear regime in log-log scales. The IV curve for SF is affected by surface barriers, whereas those under field are associated with vortex pinning [25]. The surface barriers for vortex penetration usually disappear to some characteristic field $B^* = \frac{\phi_0}{2\pi\xi w}$ related to Meissner contribution [25,26], being $\approx 8.8$ mT for the studied bridge. The $J_c^{SF}$ at 3 K are $\approx 1.2$ MA/cm$^2$ for B1 and 1.35 MA/cm$^2$ for B2, which corresponds to $\approx 45$ % of the depairing critical current $J_0$ [12]. The vortex pinning at $B > B^*$ relates to the microstructure and the thickness fluctuations [24]. The results of $J_c$ ($B$) on a log-log scale are summarized in Figure 4. The $J_c$ values decrease sharply with the field as manifested when approximating the data as a power-law with $J_c \propto B^{-\alpha}$. The $\alpha$ exponent at low fields (above Meissner contribution) is $\approx 1$, and increases to $\approx 1.4$ above $\approx 0.05$ T. The fast drop in $J_c$ ($B$) is characteristic of weak pinning, as is expected from $d << \lambda$ and smooth surfaces with low thickness fluctuations. Indeed, thin films with strong pinning usually display $\alpha$ smaller than 0.5 [26].

Figure 5a shows typical IV curves at 3 K with $\mathbf{B} \perp S$ going from the zero resistance to the normal state for B2. Similar curves with slight variations in the instability point were obtained for B1 (not shown). Inset Fig. 5a displays a close look of the transition for the fields displayed in the main panel. The evolution of the voltage with the current displays distinct regions including zero resistance, flux-flow and a switch to the normal state associated with the LO instability. This phenomenon occurs when the vortex velocity is such that the time of a vortex to move to over a distance $\approx 2\xi$ is similar to the $\tau_e$. The delocalization of quasiparticles reduces core size, and the viscosity of the flow decreases [16]. At this point, the resistivity switches to another state close to the normal one. The LO instability occurs at characteristic voltages and currents called $V_{LO}$ and $I_{LO}$, respectively.

The analysis of the LO instability is performed at $B < 1$ T, range in which the transition takes place with and abrupt jump, and therefore $V_{LO}$ and $I_{LO}$ can be extracted with small error. The value of the velocity at the instability ($v_{LO}$) relates to the inelastic relaxation time $\tau_e$ as:

$$v_{LO}^2 = 1.31 \frac{D}{\tau_e}\sqrt{1 - T/T_c}. \qquad [3]$$

The value of $v_{LO}$ depends on $V_{LO}$ as $v_{LO} = V_{LO}/(BL)$. Equation [3] is valid when the quasiparticle distribution is not confined to the vortex lattice's unit cell and becomes independent of the applied magnetic field [16]. The uniformity in the distribution of quasiparticles is reached when the product $v_{LO}\tau_e$ is higher than the intervortex distance $a = \sqrt{2\Phi_0/\sqrt{3}B}$. However, due to self-heating effects the $v_{LO}$ ($B$) $\approx$ constant regime is experimentally not observed [27,28,29].

Figure 5b shows the summary of $v_{LO}$ as a function of $B$ at 3 K for B1 and B2. Both samples display qualitatively similar behavior. $v_{LO}$ values at low fields are up to $\approx 1300$ m/s and decay as the magnetic field increases. For $B > 0.1$ T, $v_{LO}(B)$ decreases as $1/\sqrt{B}$ (see dotted lines). The shift from the $1/\sqrt{B}$ dependence at small fields is usually related to a low density of vortices and pinning inducing disorder at the vortex lattice (deviations from an Abrikosov vortex lattice) [25]. The differences in the absolute values of $v_{LO}$ between B1 and B2 may be related to local character of the instability due to geometrical imperfections in the edges of the bridges [30]. The temperature dependence of $v_{LO}$ at $B = 0.15$ T and $B = 0.49$ T with $T > 3$ K for B1 is depicted in Figure 4c. The $v_{LO}$ values drop with temperature as predicted by the equation [3]. However, the $v_{LO}$ $(B) \approx constant$ regime is not observed.

The drop in $v_{LO}$ as the magnetic field increases may be associated with an inhomogeneous distribution of quasiparticles at low magnetic fields [31] and with self-heating effects at magnetic fields in which the homogeneity is reached [27,28,29]. The data displayed in Fig. 5b do not fit, in any range of magnetic field, the model introduced by S. Doettinger et al. [31] for inhomogeneous distribution of quasiparticles (not shown). The self-heating effect is associated with finite heat removal from the substrate and it was found theoretically [27] and experimentally [28,29]. The model of Bezuglyj and Shklovskij (BS) [27] considers homogenous distribution of quasiparticles and predicts:

$$v_{LO} \propto h(1 - T/T_c)^{1/4} B^{-1/2}, \qquad [4]$$

with $h$ a thermal exchange coefficient between the film and the substrate. This parameter is associated with a transition magnetic field $B_t$ under which there is a uniform distribution of quasiparticles as:

$$B_t = \frac{0.374 e h \tau}{k_B \sigma_n d}, \qquad [5]$$

where $e$ is the electronic charge, $k_B$ the Boltzman constant, and $\sigma_n$ the normal conductivity. For $B > B_t$, dissipation in the flux-flow increases the electronic temperature and thermal effects govern the LO instability. BS proposed an scaling law for the electric field strength $E^*$ and the current density $j^*$ at the instability point:

$$\frac{E^*}{E_0} = (1 - z(b)) \left(\frac{j^*}{j_0}\right)^{-1}, \qquad [6]$$

where $E^* = \frac{V}{L} E^*$ and $j^* = \frac{I}{wd}$ are parameters of the pure LO theory and $z(b) = [1 + b + (b^2 + 8b + 4)^{\frac{1}{2}}]/3(1 + 2b)$ with $b = \frac{B}{B_t}$. The parameters $E_0$ and $j_0$ are defined as $E_0 = 1.02 B_t \left(D/\tau_\varepsilon\right)^{1/2} \left(1 - \frac{T}{T_c}\right)^{\frac{1}{4}}$ and $j_0 = 2.62(\sigma_n/e)(D\tau_\varepsilon)^{-1/2} k_B T_c \left(1 - \frac{T}{T_c}\right)^{\frac{3}{4}}$. The parameter $B_t$ can be obtained considering $P^* = E^* j^* = P_{LO}^0 (1 - z(b))$ with $P_{LO}^0 = E_0 j_0$. Moreover, $\tau_e$ $(T)$ can be obtained from the relationship $\tau_e = 2.67(B_t/P_{LO})(\sigma_n/e) k_B T_c (1 - T/T_c)$.

Figure 6 shows the scaling of the data depicted in Fig. 5b using the equation [6]. The $P^*$ $(B)$ dependences for B1 and B2 display an anomalous shoulder at 0.1 T $< B <$ 0.3 T and an increasing behavior for $B > 0.4$ T (see inset Fig. 6). Considering that the BS model is valid for uniformity in the quasiparticle distribution, we fit the data at $B > 0.4$ T using $P^* = E^* j^* = P_{LO}^0 (1 - z(b))$. The $B_t$ and $P_{LO}^0$ parameters obtained from the fits are of 0.14 (0.02) T and (9.2 (0.2) x 10$^{11}$) W/m$^2$ for B1, and 0.19 (0.01) T and (1.36 (0.05) x $10^{12}$) W/m$^2$ for B2. Using these values we estimate $\tau_e$ = 45 (5) ps for B1 and $\tau_e$ = 42 (3) ps for B2. Moreover, from equation [5], we obtain $h \approx$ 0.19 W/cm$^2$ (B1) and $h \approx$ 0.27 (B2). These $h$ values are lower than the reported for Nb [29] and NbC [25] thin films but of the same order that for NbN on polished fused quartz [32]. Using $\tau_{e,} \approx$ 45 ps and $D$ = 0.33 cm$^2$/s, we estimate a diffusion length of exited quasiparticles $l_e = (D\tau)^{1/2} \approx$ 39 nm. This value corresponds to an intervortex center space with $B \approx$ 1.3 T, which is higher than the values of field used for the analyses (0.4 T < $B$ < 0.85 T) and much larger than the extracted $B_t$. Considering that the cores have a finite diameter $\approx 2\xi$, the magnetic field in which the distance between core boundaries is $\approx$ 39 nm reduces below 1 T.

The results obtained in this work show that nanocrystalline $\gamma$-Mo$_2$N thin films display an excellent performance for application in SNSPD. The films are homogenous and preserve superconductivity for thicknesses of a few nanometers. Moreover, the $T_c$ value increases if surface passivation is avoided. Based on chemical compatibility, the surface may be protected using materials with active optical properties [33,34]. The films present upper critical field above 10 T and very low vortex pinning. From the LO instability, we determine a $\tau_e \approx$ 45 ps, which is similar to amorphous MoSi ($\approx$76 ps) [3] and NbN (12 ps- 88 ps) [35] and smaller than that experimentally determined for WSi ($\approx$193 ps) [4].

## 4. Conclusions

In summary, we have demonstrated that sandwich structures avoid dead layers and enhance the superconducting critical temperature of ultra-thin $\gamma$-Mo$_2$N films. The films are extremely flat and display weak vortex pinning. Moreover the self-field $J_c$ at $T$ = 3 K is close to the 45 % of $J_0$ (0). Using the LO instability, we estimate a relaxation time of quasiparticles of $\approx$ 45 ps for a film with a thickness of 6.4 nm. This value is close to the previously reported for materials that display high performance for applications in single-photon detectors.

### Acknowledgment

This work was partially supported by the ANPCYT (PICT 2018- 01597 and 2018-01597), U. N. de Cuyo 06/C576 and CONICET PIP 2015-0100575CO. SB and NH are members of the Instituto de Nanociencia y Nanotecnología INN (CNEA-CONICET).

### Declaration of competing interest

The authors declare that they have no known competing financial interestsor personal relationships that could have appeared to influencethe work reported in this paper.

### Author Contributions

J. A. H., M. G. and N.H. growth the films and performed the transport measurements. N.H. performed the XRD patterns. S.B. performed the XPS spectra. All authors contributed to the analysis of the results and drafting the manuscript.

List of Figures:

Figure 1. XPS Mo3d spectra of the surface of a $Mo_2N$ thin films with (panel *a)* and without (panel *b)* an AlN capping layer.

Figure 2. *a-b)* XRR and corresponding fits for single $Mo_2N$ films and sandwich structures with AlN.*c)* Temperature dependence of the normalized resistivity ($R/R^{10K}$) for the different samples. *d)* Summary for the $T_c$. For single films the $T_c$ is plotted for the nominal and the effective thickness obtained from XRR.

Figure 3. Temperature dependence of the upper critical field ($H_{c2}$) with the magnetic field applied parallel and perpendicular to the surface. Fits using equations [1] and [2] are included. Inset: Typical curves of normalized resistance ($R/R^{10\ K}$) with the magnetic field applied parallel to the surface.

Figure 4. Magnetic field dependence of $J_c$ at T = 3 K with **B** $\perp$ S for two different bridges of $[Mo_2N/AlN]_{6.4}$. Inset: Characteristic IV curves at self-field and under applied magnetic field. The criterion for $J_c$ is indicated.

Figure 5. *a)* Characteristic IV curves at 3 K and several representative magnetic fields for B1. Inset shows the different regimes at the IV curves in log-log scales at $B$ = 0.1 T. *b)* Dependence of the critical velocity on the magnetic field in the LO instability at 3 K for B1 and B2. *c)* Temperature dependence of the critical velocity at $B$ = 0.16 T and $B$ = 0.49 T for B2.

Figure 6. Comparison between $\left(\dfrac{\frac{E^*}{E_0}}{1-z(b)}, \left(\dfrac{j^*}{j_0}\right)^{-1}\right)$ a lineal scaling considering $y \propto x$ with x = 1 / $j^*$ (equation 6). Inset: $P^*$ ($B$). The two parameters $B_t$ and $P_{LO}^0$ are deduced from these fits for $B > 0.4$ T.

Figure 1.

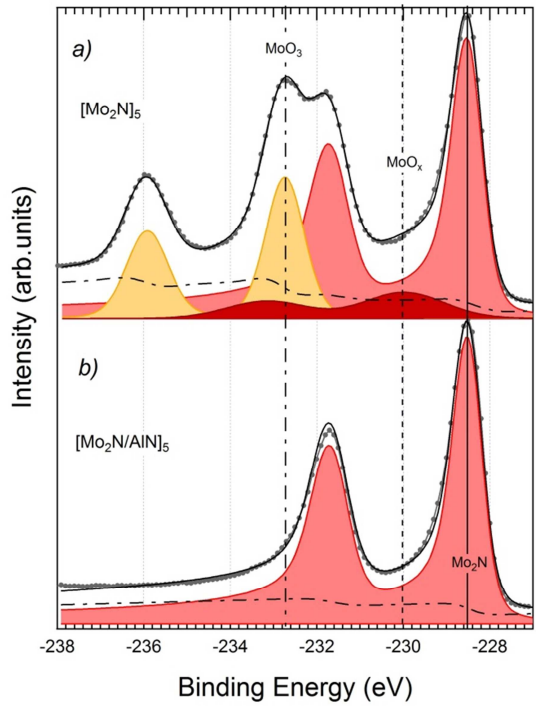

Figure 2.

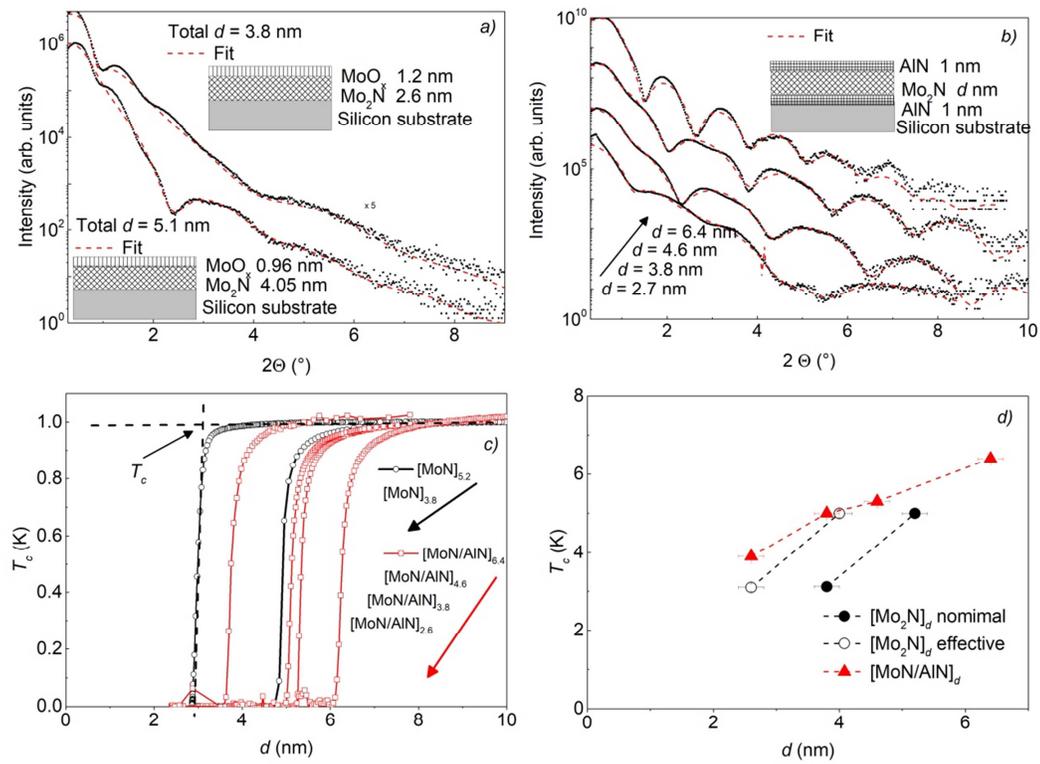

Figure 3.

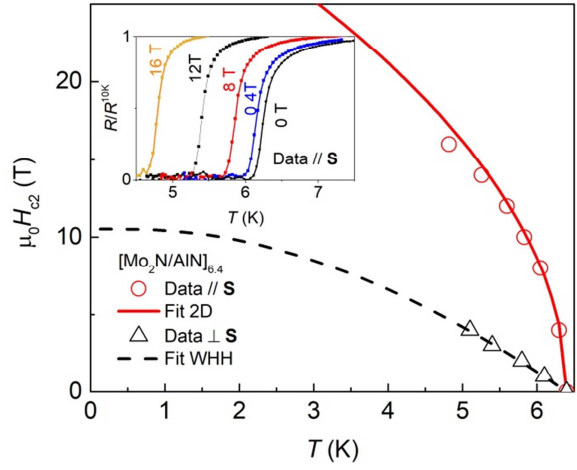

Figure 4.

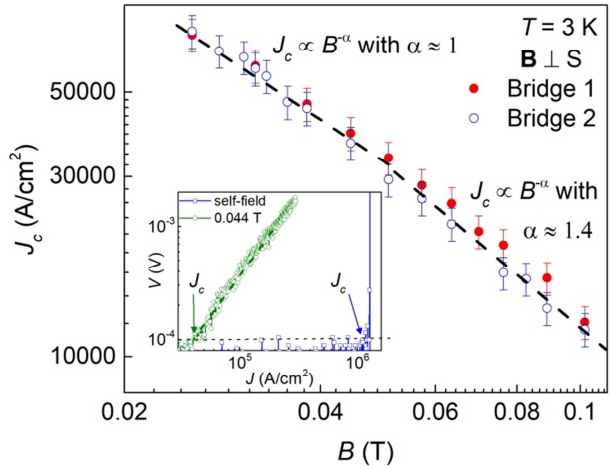

Figure 5.

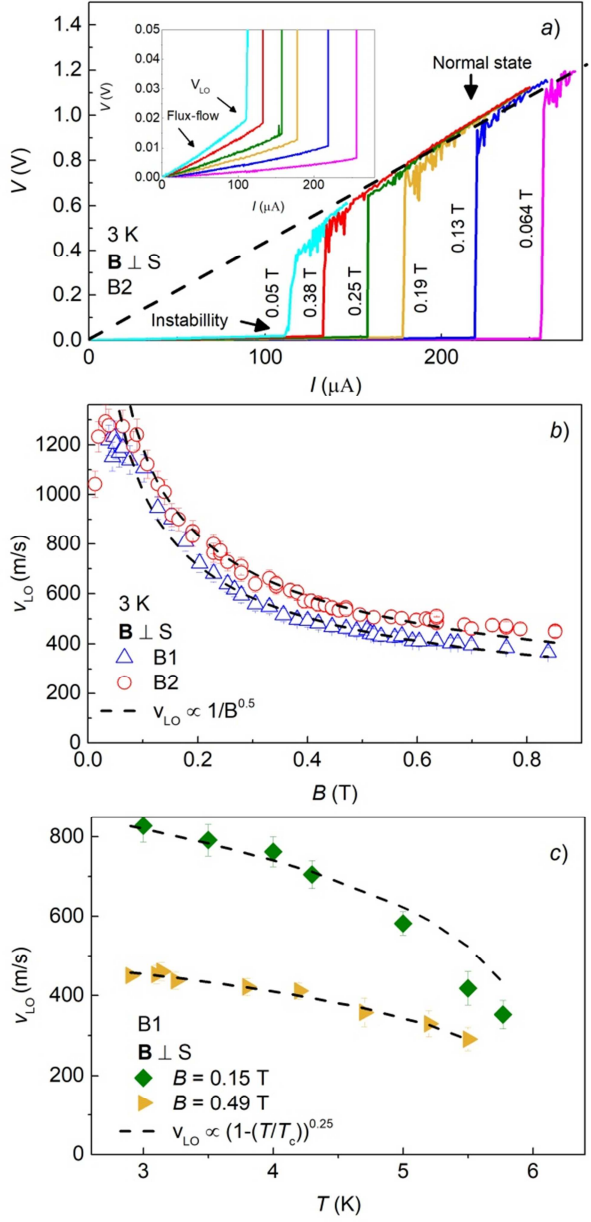

Figure 6.

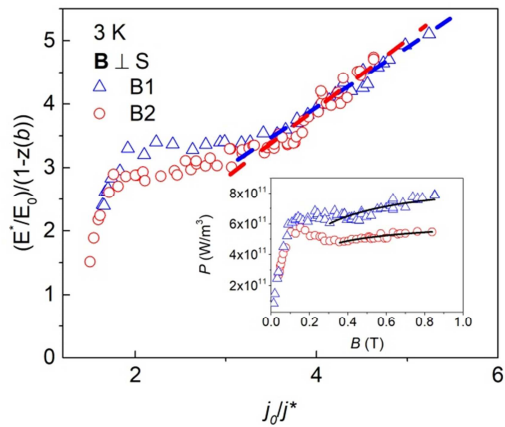